\documentstyle[prl,aps]{revtex}
\begin{document}
\twocolumn 
%\draft
%\preprint{HEP/123-qed}
\title{Absence of a Magnetic Field Induced Metal-Insulator Transition in Kondo 
Insulators} 
\author{Herv\'e M. Carruzzo and Clare C. Yu} 
\address{
Department of Physics, University of California, Irvine, Irvine, California
92717 }
\date{\today}
\maketitle
\begin{abstract}
We investigate whether or not Kondo insulators undergo a magnetic field
induced metal-insulator transition in one dimension at half filling 
using both a density matrix formulation of the numerical
renormalization group and a new analytical approach. 
Contrary to expectations, the
quasiparticle gap never vanishes at any field
and no metal-insulator transition is found. We discuss generalizing
our result to three dimensions and to the asymmetric Anderson lattice.
Our result is consistent with recent experiments.
\end{abstract}

\narrowtext 

In a strong magnetic field an ordinary narrow gap
semiconductor becomes metallic
when the field is comparable to the size of the gap.
Is this also true for the class of rare earth
compounds known as Kondo insulators, whose 
insulating gap is due to interactions
between conduction electrons and localized f-electrons \cite{Fisk}?
Simple arguments
\cite{millis,carlos,varma}
suggest that such a metal-insulator transition (MIT) should occur 
when the applied magnetic field $h$ exceeds the exchange coupling $J$
but is less than the conduction electron bandwidth. In this case,
one expects the f-spins to be completely polarized and
decoupled from the conduction electrons, leaving an incompletely polarized
band of metallic conduction electrons. This scenario
is supported by mean field calculations such as
those with slave bosons \cite{millis,hmc}. 
However, here we show that such expectations are not correct when 
spin fluctuations are taken into account. 

We have studied the behavior of the insulating 
gap in a one-dimensional Kondo insulator as a function of magnetic field.
Using the density matrix formulation of the numerical renormalization 
group\cite{white1,white2} as well as a new analytic approach, 
we find that the gap, while greatly
reduced by the applied magnetic field, never vanishes at any 
value of $h$. This unexpected result is due to
spin flip scattering across the Fermi surface which removes
degeneracies and keeps the gap open. We 
argue that this result also holds in  
three dimensions and for 
the asymmetric Anderson lattice where there is no particle-hole
symmetry. Our results are consistent with 
recent resistivity and Hall measurements on Ce$_3$Bi$_4$Pt$_3$ 
in pulsed fields up to 61 T
which do not see the signatures expected for a transition
from a semiconductor to a good metal \cite{boebinger}. 
This material acts like
a semimetal or a dirty metal at the lowest temperatures ($T<5$ K)
and highest fields ($h>50$ T), even though the field exceeds the
experimentally deduced value of the gap 
($\Delta\sim 37$ K) \cite{schlesinger}.

The 1D Kondo lattice has
spin-1/2 conduction electrons that hop from site
to site with an on-site spin exchange between a local f-electron
and the conduction electron on that site.
Thus the Hamiltonian is
\begin{eqnarray}
\nonumber
H=-t\sum_{i,\sigma }&&\left( c_{i\sigma }^{\dagger}c_{i+1\sigma }
+h.c.\right)
\\  
&&+\sum_iJ\vec S_{if}\cdot \vec S_{ic}  
-h\sum_i\left( S_{ic}^z+S_{if}^z  
\right) ,
\end{eqnarray}
where the conduction electron spin density on site i is $\vec S_{ic}
=\sum_{\alpha\beta}c_{i\alpha}^{\dagger}(\vec\sigma /2)_{\alpha\beta}  
c_{i\beta}$, the f-electron spin density is $\vec S_{if}  
=\sum_{\alpha\beta}f_{i\alpha}^{\dagger}(\vec\sigma /2)_{\alpha\beta} 
f_{i\beta}$, $\vec\sigma_{\alpha\beta}$ are Pauli matrices and h
is the external magnetic field, chosen along the z-axis. We set the 
hopping matrix element $t=1$ and choose
$J$ to favor antiferromagnetic ordering ($J>$0). 
To study the Kondo insulator, we restrict ourselves to 
half filling where the total number of conduction electrons $N$ equals
the number of sites $L$.

We consider a system to be insulating when the quasiparticle gap  
$\Delta_{qp}\equiv\mu_{N+1}-\mu_{N}\not= 0$. Here  
$\mu_{N}=E_{N}-E_{N-1}$ is the chemical potential and 
$E_{N}$ is the ground state energy with N electrons.
We have calculated $\Delta_{qp}$
using both the density matrix renormalization
group (DMRG) algorithm \cite{white1,white2,yu} and a new 
analytic technique. We first
describe the results of our numerical work.

The DMRG approach is a real space technique which has proven
to be remarkably accurate for the Kondo lattice \cite{yu}.
We used the finite system method \cite{white2} with open
boundary conditions in which there is no hopping past the
ends of the chain. We studied lattices of size
$L$=6, 8, 16, and 32, keeping up to
100 states. The energies were extremely
accurate for $J\gg t$, with typical truncation errors of order $10^{-9}$
%for $J$=10. For $J\stackrel{\textstyle <}{\sim}t$, the f-spin
for $J$=10. For $J{{ }_<\atop { }^{\sim}}t$, the f-spin
degrees of freedom lead to a large number of nearly degenerate
energy levels. As a result, the
accuracy was significantly reduced, with truncation errors of
order $10^{-4}$ for $J$=0.75. 

To determine $\Delta_{qp}$, we calculated
the ground state energies with $N=L$ and $N=L+1$ conduction electrons
\cite{phsym} for all possible values of the total $S^{z}$ with $h=0$. 
Since the magnetic field term in the Hamiltonian commutes with
the Hamiltonian, we can find the energies in
the presence of a field by adding $-hS^{z}_{tot}$ to the zero field
energies. We then use the lowest energies
($E_{N}^{min}(S^{z}_{tot})$ and $E_{N+1}^{min}(S^{z}_{tot})$)
as a function of $S^{z}_{tot}$ to calculate $\Delta_{qp}$.
In the presence of a field, finite size effects do not limit
the size of the gap because the field can tune the levels
$E_{N}$ and $E_{N+1}$
to be infinitesimally close. For example, for free electrons
on an 8 site lattice, the minimum $\Delta_{qp}$ is zero
to within the roundoff error of $10^{-15}$. In Fig. 1 we show
the quasiparticle gap as a function of $h$. The sawtooth
oscillations occur as the ground states, $E_{N}$ and $E_{N+1}$,
change their polarizations with increasing $h$.

Notice that at high fields, the gap is proportional to $h$. 
This occurs when the spins 
of the half-filled lattice 
are completely polarized by the field. 
To calculate $E_{N+1}$, we add
an electron whose spin is opposite to the field. The energy required
to do this increases linearly in $h$, and hence $\Delta_{qp}$ is
proportional to $h$. This can also be understood in terms of splitting
the up(+) and down($-$) spin conduction electron bands. For 
fully polarized f-spins, the conduction electron energies are 
$E_{\pm}=-2tcos(k)\pm (J/4 -h/2)$. The two bands are   
completely separated at a field $h_{s}=J/2+4t$ and so, 
for $h>h_{s}$, $\Delta_{qp}\sim h$. Our expression for $h_s$ is  
only an estimate since, as we shall see, the f-spins are not fully polarized
until the bands separate.

For a given value of $J$, we can find the minimum quasiparticle
gap, $\Delta_{qp}^{min}$, as a function of $h$ from plots similar to Fig. 1.
Roughly speaking, the minimum occurs at 
the largest field in which the up and
down spin bands are not completely separate. This
field, $h_{min}$, is of order the
conduction electron bandwidth, i.e., $h_{min}\sim h_{s}\sim 4t$.
Notice that the minimum does not occur at $h\sim J$ or at
$h\sim\Delta_{qp}$ as one might expect.
Fig. \ref{fig:mingap} shows $\Delta_{qp}^{min}$    
as a function of $J$. Fitting the curves
for $J\leq 2$ to $\Delta_{qp}^{min}\sim J^{b}$, we 
find that $b$ increases from
2.05 for $L$=6 to 2.16 for $L$=32. The power law form indicates that the
gap does not vanish at any finite value of $J$.

%To understand why the gap remains open, note that the f-spins are
To see why the gap remains open, note that the f-spins are
not completely polarized for $h\leq h_{s}$. This allows 
spin flip scattering across the Fermi surface
between the conduction electrons
and the f-spins. It removes degeneracies at the Fermi energy 
and prevents the quasiparticle gap from closing. 
The inset of Fig. \ref{fig:gapvsH} shows the conduction band structure
for a nonzero $h$. Note that the Fermi points satisfy 
$k_F^{R+}-k_F^{L-}=\pi$ and $k_F^{R-}-k_F^{L+}=\pi$ for $h\leq h_{s}$.
Thus scattering processes with momentum
transfer $2k_F=\pi$ will be important.
($k_F$ is the Fermi wavevector in zero field.)

We have verified this scenario using a new analytic approach
in which we keep two types of scattering processes: 
(1) those with no momentum transfer
and no spin flip and (2) those with spin flip scattering 
with $\pi$ momentum transfer. We call these S1 and S2, respectively.
These processes are not restricted to the Fermi surface.

We use the Popov-Fedotov functional integral approach to enforce the 
constraint of 1 f-electron per site\cite{popov}.  
Our strategy is to start from the partition function $\exp(-\beta H)$ and 
integrate out the f degrees of freedom self-consistently to obtain an 
effective action for the conduction electrons. This yields the
conduction electron propagator whose poles determine the $T$=0 excitation 
spectrum. The gap is  
the smallest excitation energy. We find a nonzero gap for
all values of the field.

We adopt a basis of Right and Left variables for the conduction   
and f-states. If $\psi$ is a $c$ or $f$ operator, then 
\begin{mathletters} 
\begin{eqnarray}
\psi_{R\alpha}(k,\omega )&\equiv &
\left\{\matrix{\psi_{\alpha}(k,\omega ) & k\geq 0\cr
0&\hbox{\rm otherwise}\cr}\right. \\
\psi_{L\alpha}(k,\omega )&\equiv &
\left\{\matrix{\psi_{\alpha}(k-\pi,\omega ) & k>0\cr
0&\hbox{\rm otherwise}\cr}\right.
\end{eqnarray}
\end{mathletters}
Thus, $2k_F$ momentum scattering processes link R and L variables, while
zero momentum terms involve R or L states only. 
We write the partition function, in a Fourier representation, 
as follows:
\begin{mathletters} 
\begin{eqnarray}
{\cal Z}&\sim &\int {\cal D}\bar c{\cal D}c{\cal D}\bar f{\cal D}f 
e^{-S_{oc}-S_{of}-S_{fc}} \\
S_{oc}&=&{1\over\beta}\sum_{\omega}\int_0^{\pi}{dk\over 2\pi}  
\bar c(k,\omega )D_oc(k,\omega )  \\  
S_{of}&=&{1\over\beta}\sum_{\omega}\int_0^{\pi}{dk\over 2\pi}
\bar f(k,\omega )F_of(k,\omega )  \\ 
S_{fc}&=&{1\over\beta^2}\sum_{\omega_1\omega_2}\int_0^{\pi}{dk_1dk_2\over
(2\pi )^2}\bar f(k_1,\omega_1)C_{\Delta k}^{\Delta\omega}f(k_2,\omega_2) 
\end{eqnarray} 
\end{mathletters}
\noindent where $c$ and
$f$ are now vectors of Grassman variables of the form
$\bar\psi =(\bar\psi_{R+},\bar\psi_{L+},\bar\psi_{R-},
\bar\psi_{L-})$. $\omega$ is a fermion Matsubara frequency, 
i.e., $\omega=2\pi(n+1/2)/\beta$.
$\Delta k=k_1$$-k_2$, $\Delta\omega =\omega_1-\omega_2$, and $\beta$ is
the inverse temperature. 
$S_{oc}$ 
and $S_{of}$ are the actions for the free conduction electrons 
and the f-states 
respectively, and $D_o^{-1}$ and $F_o^{-1}$ are the corresponding (imaginary 
time) propagators: 
\begin{eqnarray}
\nonumber 
D_o&\equiv &\left[\matrix{i\omega -h-2t\cos (k)\sigma^z & 0 \cr
0 & i\omega +h-2t\cos (k)\sigma^z \cr }\right] \\
\nonumber 
F_o&\equiv &\left[\matrix{i\omega -i\pi / 2\beta -h &0\cr
0& i\omega -i\pi /2\beta +h \cr}\right]   
\end{eqnarray} 
$S_{fc}$ contains the coupling between the conduction
electrons and the f-states:  
\begin{eqnarray} 
C_{\Delta k}^{\Delta\omega}&\equiv & 
\left[\matrix{A & 2B_{-+} \cr
2B_{+-} & -A \cr}\right]  
\ B_{\alpha\beta}\equiv \left[\matrix{R_{\alpha\beta,0} & R_{\alpha\beta,\pi}
\cr R_{\alpha\beta,\pi} & R_{\alpha\beta,0} \cr }\right] \\
\nonumber 
A&\equiv &\left[\matrix{R_{++,0}-R_{--,0} & R_{++,\pi}-R_{--,\pi} \cr
R_{++,\pi}-R_{--,\pi} & R_{++,0}-R_{--,0} \cr }\right]  
\end{eqnarray} 
$D_o$, $F_o$ and $C_{\Delta k}^{\Delta\omega}$ are 4x4 matrices while A and 
$B_{\alpha\beta}$ are 2x2 matrices. $R_{\alpha\beta,0}$ 
represents scattering processes without spin flip (``0''):  
\begin{eqnarray}
\nonumber 
&&R_{\alpha\beta ,0}={J\over 4\beta}\sum_{\omega}\int_0^{\pi}
{dk_3dk_4\over (2\pi )^2}\{  
\bar c_{R\alpha}(k_3,\omega )c_{R\beta}(k_4,\omega +\Delta\omega) \\  
&&+\bar c_{L\alpha}(k_3,\omega )c_{L\beta}(k_4,\omega +\Delta\omega ) 
\}\bar\delta (k_4-k_3-\Delta k)  
\end{eqnarray}  
and $R_{\alpha\beta,\pi }$ describe spin flip scattering across the Fermi
surface ($\pi$):  
\begin{eqnarray}
\nonumber
&&R_{\alpha\beta ,\pi}={J\over 4\beta}\sum_{\omega}\int_0^{\pi}
{dk_3dk_4\over (2\pi )^2}\{ 
\bar c_{R\alpha}(k_3,\omega )c_{L\beta}(k_4,\omega +\Delta\omega )\\  
&&+\bar c_{L\alpha}(k_3,\omega )c_{R\beta}(k_4,\omega +\Delta\omega ) 
\}\bar\delta (k_4-k_3-\Delta k-\pi )  
\end{eqnarray}  
\noindent where $\bar\delta$ enforces momentum conservation mod$2\pi$,
i.e., umklapp scattering is included. 

Since we are interested in the $c$ fields, we  
integrate out the $f$ fields. This leaves an effective action 
$S_{eff}$ for the
conduction electrons:  
\begin{eqnarray}
{\cal Z}\sim\int \ e^{-S_{eff}}=
\int \ e^{-S_{oc}+\ln (\hbox{\rm det}(F_o+C))}   
\end{eqnarray}
An expansion of the ln(det..))
around $F_o$ would correspond to perturbation theory in terms of the bare
$F_o^{-1}$ propagator. This is not a good starting point since it ignores
the effect of the conduction electrons on the f-states. 
Let $\langle ...\rangle$ denote ${1\over {\cal Z}}$
$\times\int {\cal D}\bar f{\cal D}f\exp\{-S_{eff}\}(...)$. 
We replace $F_o^{-1}$ by its ``dressed'' version, $(F_o+\langle C\rangle )
^{-1}$, which includes the average effect of the conduction electrons 
($\langle C\rangle$) on the f-states themselves\cite{graphs}. 
$\langle C\rangle$ acts like an f-state self-energy.
Thus we write    
\begin{eqnarray} 
\nonumber
&&\ln (\hbox{\rm det}(F_o+C))\rightarrow \ln (\hbox{\rm det}(F_o+
\langle C\rangle +C)) \\ 
&&\approx \hbox{\rm Tr}\ln (F_o+\langle C\rangle )\ 
+\ \hbox{\rm Tr}((F_o+\langle C\rangle )^{-1}C) 
\label{trln} 
\end{eqnarray}   
$\hbox{\rm Tr}\ln (F_o+\langle C\rangle )$ is a constant
which we will ignore. In the matrix $C^{\Delta\omega}_{\Delta k}$,
we set $\Delta k=0$ and $\Delta\omega=0$ which correspond
to S1 and S2 scattering processes.
The only nonzero components of $\langle C\rangle$ are the diagonal terms of
A:
\begin{equation}
C_o\equiv (\langle R\rangle_{++,0}-\langle R\rangle_{--,0})\beta
\delta (\Delta k)
\delta (\Delta\omega ) \label{etao} 
\end{equation} 
and the off-diagonal terms of B: 
\begin{equation} 
C_{\pi}\equiv \langle R\rangle_{+-,\pi}\beta\delta (\Delta k)\delta (
\Delta\omega ) \label{etapi} 
\end{equation} 
since $\langle R\rangle_{+-,\pi}=
\langle R\rangle_{-+,\pi}$, by symmetry. 

The last term in eq. (\ref{trln}), together with $S_{oc}$,
gives our approximation to $S_{eff}$:
\begin{eqnarray}
&&S_{eff}={1\over\beta}\sum_{\omega}\int_0^{\pi}{dk\over 2\pi}
\bar c(k\omega )Dc(k\omega )  \\
\nonumber
&&D=\left[\matrix{
i\omega -\tilde h -2t\cos (k)\sigma^z &
-{JC_{\pi}\over 4\lambda}\sigma^x \cr
-{JC_{\pi}\over 4\lambda}\sigma^x &
i\omega -\tilde h -2t\cos (k)\sigma^z
\cr }\right]
\end{eqnarray}
where $\lambda =\sqrt{(h/2-C_o)^2+C_{\pi}^2}$ and  
$\tilde h =h/2-J(h-2C_o )/8\lambda$.  
Analytic continuation of $D^{-1}$ to real time yields the effective
conduction electron propagator whose poles 
give the excitation spectrum which consists of four
branches. The upper two branches are 
$\epsilon_{\pm}(C_o,C_{\pi})=\sqrt{ 
(|\tilde h|\pm |\gamma |)^2+\left(C_{\pi}J/2\lambda\right)^2}$,
where $\gamma =2t\cos (k)$, and the lower two branches are 
$-\epsilon_{\pm}(C_o,C_{\pi})$.
The gap between the upper and lower branches is 
$\Delta_{qp}=C_{\pi} J/\lambda$. Thus 
$\Delta_{qp}\rightarrow 0$ 
would signal a MIT. 

To determine the gap, we need $C_{\pi}$ and $\lambda$; $\lambda$
is a function of $C_o$ and $C_{\pi}$. 
In eqs. (\ref{etao}) and (\ref{etapi}) $C_o$ and $C_{\pi}$ involve 
averages over $S_{eff}$, which is itself a 
function of these parameters. As a result, we get self-consistent
coupled equations for $C_{\pi}$ and $C_o$: 
\begin{mathletters} 
\begin{eqnarray}
C_o&=&{J\over 4}\int_0^{{\pi\over 2}}{dk\over\pi}\left\{
{\tilde h+\gamma sgn(\tilde h )\over\epsilon_+(C_o,C_{\pi})}+ 
{\tilde h -\gamma sgn(\tilde h )\over
\epsilon_-(C_o,C_{\pi})}\right\}
\label{eta01} \\
\lambda &=&\left({J\over 4}\right)^2\int_0^{{\pi\over 2}}
{dk\over\pi}\left\{ 
{1\over\epsilon_+(C_o,C_{\pi})}+
{1\over\epsilon_-(C_o,C_{\pi})}\right\}    
\label{etapi1} 
\end{eqnarray}  
\end{mathletters} 
Inspection of eq. (\ref{eta01}) shows that $|C_o|<J/4$, which
indicates that $\lambda$ will not diverge due to $C_o$ diverging.
(If $\lambda$ diverges, the gap goes to zero.)
To find $C_{\pi}$, note that as long as the two lower bands are
not completely separated in energy, i.e., $|\tilde h |\leq 2t$,
the right hand side of 
eq. (\ref{etapi1}) diverges logarithmically as $C_{\pi}\rightarrow 0$.
This singularity guarantees that
the self-consistent equations can always be satisfied by a finite
value of $C_{\pi}$ 
for $|\tilde h|\leq 2t$. Thus the gap {\it never
vanishes}. (For $|\tilde h|> 2t$, the gap grows linearly with h as discussed
earlier.) This conclusion does not rely on
the precise shape of the band, nor on $\pi$ momentum transfer for  
the scattering processes. 

We can solve eqs. (\ref{eta01}) and (\ref{etapi1}) numerically
for the gap.
As shown in Figs. \ref{fig:gapvsH} and \ref{fig:mingap}, there is
good agreement between our analytic (adapted for finite size) and 
DMRG approaches which
indicates that our analytic calculation contains the essential physics.
In Fig. \ref{fig:mingap}
the analytic results are smaller since they include only a subset
of all possible scattering processes. Note that the analytic
results are for periodic boundary conditions.

Using eqs. (\ref{eta01}) and (\ref{etapi1}), we find that
$\Delta_{qp}^{min}$ 
for an infinite lattice becomes very small as $J\rightarrow 0$
(see in the inset of Fig. \ref{fig:mingap}) \cite{extrap}.
For a linear conduction band ($\gamma=2t|1-2k/\pi|$)
in the limit of small $J$,
\begin{equation}
\Delta_{qp}^{min} ={g_1\over J}\times e^{-g_2/J^2}
\label{gap}
\end{equation} 
where $g_1$, $g_2$ are real positive constants. 
The factor of $J^{-2}$ in the exponent
shows that pairs of spin flips dominate, e.g., an f-spin,
which points up in the field, flips from up to down and
back to up.
This functional form also fits the gap
%resulting from a cosine conduction band ($\gamma=2t\cos(k)$) as shown in
for a cosine conduction band ($\gamma=2t\cos(k)$) as shown in
Fig. \ref{fig:mingap}, though we could test the fit only over
a rather limited range of $J$ since the extremely small values of
$\Delta_{qp}^{min}$ for $J<1.8$ were obscured by numerical noise.

We argue that there is also no field induced MIT 
for the half-filled asymmetric Anderson lattice, which lacks
particle-hole symmetry. 
The external field pushes the f-band, which has spins
aligned along the field, away from the Fermi energy and
into the Kondo regime, where charge fluctuations are reduced.
This can be shown using the Schrieffer-Wolff transformation.
DMRG calculations on small lattices ($L\leq 8$) also
do not find a MIT\cite{Mariana}.

We believe that spin flip scattering will also keep the gap open
in three dimensions. The major
difference in 3D is the formation of Landau ``tubes''. We conjecture that the
associated van Hove singularities will actually lead to enhanced 
scattering and hence to a larger $\Delta_{qp}^{min}$
than in the 1D case.

To summarize, 1D Kondo insulators do not
undergo a magnetic field induced
MIT due to strong spin  
flip scattering across the Fermi surface. 
Using a new analytic
approach, we find that for an infinite lattice, the
quasiparticle gap depends exponentially on $J^{-2}$
for small $J$.
We suggest that the absence of a field induced MIT  
applies to the 3D case as well as the asymmetric Anderson model. 
The fact that a large field can significantly reduce but not
close the insulating gap is consistent with 
experiments which find that Ce$_{3}$Bi$_{4}$Pt$_{3}$ acts
like a semimetal in high fields \cite{boebinger}.

We thank
Greg Boebinger, 
Zack Fisk, 
Mariana Guerrero, 
Jon Lawrence, 
Andy Millis,  
Steve White
and Sue Coppersmith
for helpful discussions. We also thank Greg Boebinger for
sending us his data before publication.
This work was
supported in part by ONR Grant No. N000014-91-J-1502  
and Los Alamos National Laboratory. C.C.Y. is an
Alfred P. Sloan Research Fellow.

\begin{figure}
\caption{Quasiparticle gap versus $h$ for $J=2$ with $L=32$. 
The thick solid line is the DMRG result with open boundary
conditions, while the thin line is the analytic result
with periodic boundary conditions.
Inset: Typical conduction electron band structure for 
$J=0$, $t=1$ and $h=2$.}
\label{fig:gapvsH} 
\end{figure} 

\begin{figure}
\caption{Minimum quasiparticle gap versus $J$ for $L$ = 8, 16, and 32.
Solid symbols are DMRG results; open symbols are analytic
results. Solid lines are guides for the eye.
Inset: Minimum quasiparticle gap versus $J$ for an infinite lattice.
The solid line is a fit to eq. (\protect\ref{gap}) with $g_{1}=12.13$ and
$g_{2}=57.97$.}
\label{fig:mingap} 
\end{figure} 


\begin{references}
\bibitem{graphs} A diagrammatic analysis shows this procedure to be
equivalent to resumming all contributions to the conduction and f-state
self-energies with 0 or $\pi$ momentum transfer and no extra internal
$k$ and $\omega$ sums.
\bibitem{extrap} $\Delta_{qp}^{min}$ calculated analytically for $L=\infty$
is several orders of magnitude smaller than the result of extrapolating
the DMRG curves to $L=\infty$ by plotting the data versus
$1/L$ for a given value of $J$. Extrapolation of
finite size data neglects the fact that Fermi surface scattering from
the large density of states at the bottom of the conduction
band when $h\sim h_s$ leads to a larger gap than when
$h<h_s$. (The density of states becomes large as $L$ increases.)
\end{references}
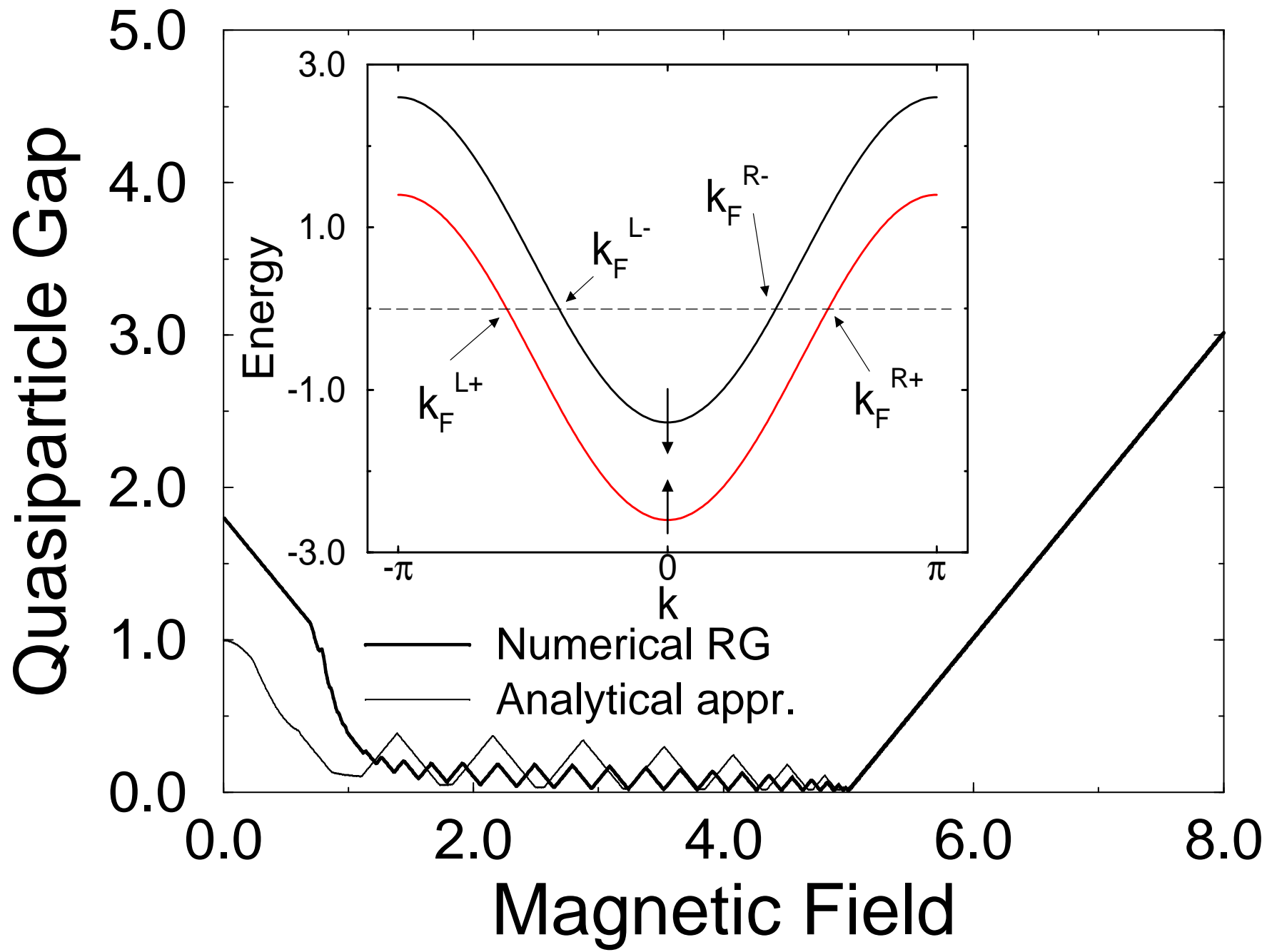
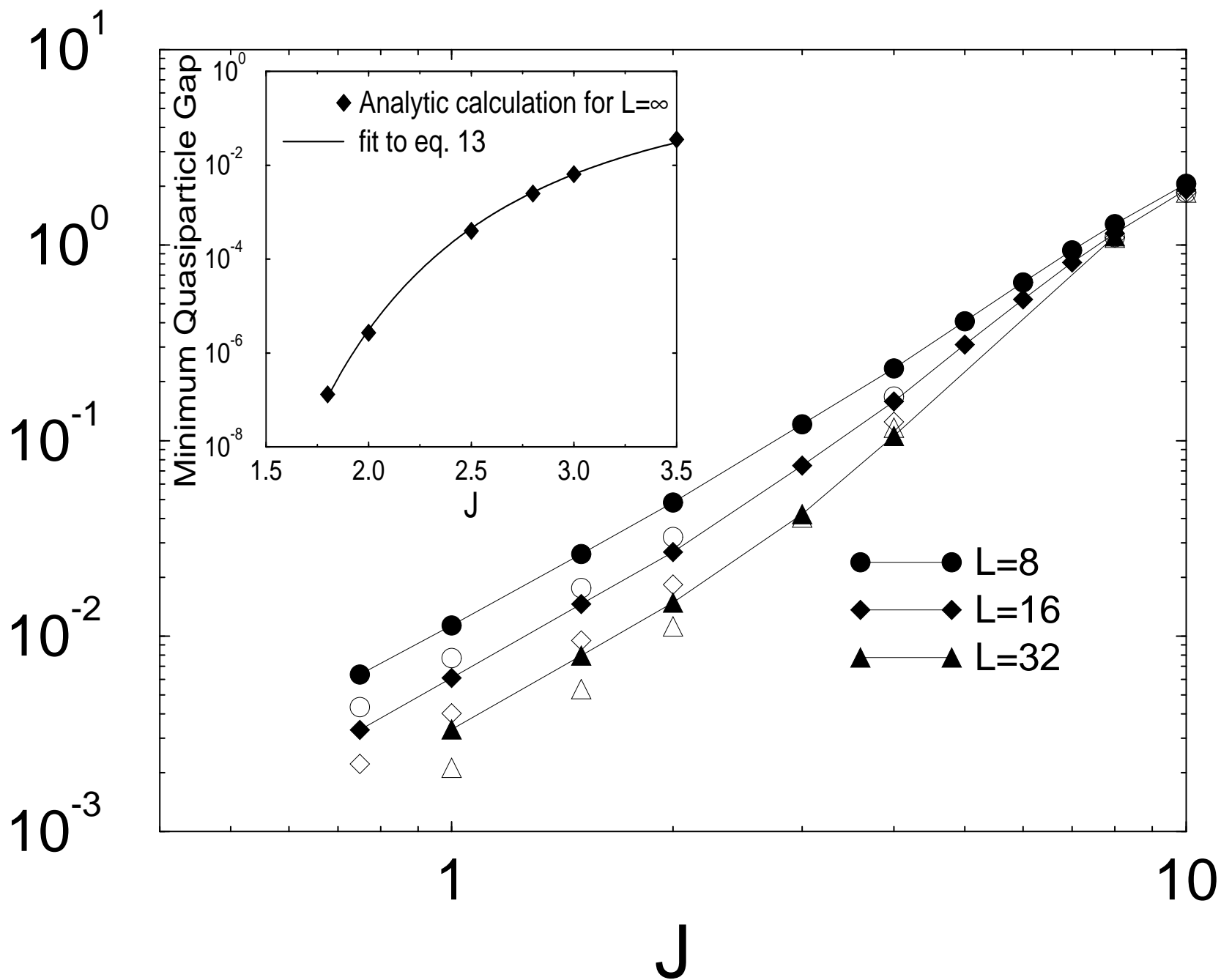
\end{document}